\documentstyle[12pt,epsf]{article}

\begin{document}

\newcommand{\balpha}{{\mbox{\boldmath$\alpha$}}}
\newcommand{\bmu}{{\mbox{\boldmath$\mu$}}}
\newcommand{\bnu}{{\mbox{\boldmath$\nu$}}}
\newcommand{\be}{\begin{eqnarray}}
\newcommand{\ee}{\end{eqnarray}}
\newcommand{\la}{\langle}
\newcommand{\ra}{\rangle}
\newcommand{\bfx}{{\bf x}}
\newcommand{\bfy}{{\bf y}}
\newcommand{\az}{\alpha Z}
\newcommand{\bfz}{{\bf z}}
\newcommand{\bfn}{{\bf n}}

\begin{center}

{\Large \bf
Quantum electrodynamics of heavy ions and atoms }
\end{center}
\begin{center}
{\large V.M. Shabaev, A.N. Artemyev, D.A. Glazov,
 I.I. Tupitsyn, V.A. Volotka,\footnote{Present address:  
Institut f\"ur Theoretische Physik, TU Dresden,
Mommsenstra{\ss}e 13, D-01062 Dresden, Germany} 
and V.A. Yerokhin}
\end{center}
$$
\,
$$
\begin{center}
{\it Department of Physics, St.Petersburg State University},\\
 {\it Oulianovskaya 1, Petrodvorets, St.Petersburg 198504, Russia}
\vspace{0.5cm}
\end{center}
PACS number(s): 12.20.-m, 12.20.Ds, 31.30.Jv, 31.30.Gs

\begin{abstract}

The present status of quantum electrodynamics (QED) theory of
heavy few-electron ions is reviewed.
The theoretical results
are compared with available experimental data.
A special attention is focused on tests of
QED at strong fields and on determination
of the fundamental constants. A recent progress on calculations
of the QED corrections to the parity nonconserving 6s-7s transition
amplitude in neutral Cs is also discussed.

\end{abstract}

\section{Introduction}

Basic principles of quantum electrodynamics (QED) were formulated
to the beginning of 1930's as the result of merging quantum mechanics
and special relativity.
This theory provided description of such low-order 
processes as emission
and absorption of photons and creation and annihilation of 
electron-positron pairs. However, application of this theory to
calculations of some higher-order effects gave infinite results. 
This problem remained unsolved till the late of 1940's, when
Lamb and Retherford discovered the $2p_{1/2} - 2s$ splitting in hydrogen,
which is presently known as the Lamb shift.
This discovery stimulated 
theorists to complete the creation of QED since it was believed that this
splitting is of QED origin. 

First evaluation of the Lamb shift was given by Bethe. 
The rigorous QED formalism was developed by Dyson, Feynman, 
Schwinger and Tomonaga. They found that all divergences can be removed
from the theory by the  renormalization procedure.
All calculations in QED are
based on the perturbation theory in a small parameter, which 
is the fine structure constant $\alpha \approx 1/137$. 
Individual terms of the perturbation
series are conveniently represented by Feynman diagrams.
For instance, the lowest-order contribution
to the Lamb shift in a H-like atom is determined by so-called
self-energy and vacuum-polarization diagrams, presented in Fig. 1.



\begin{figure} \caption{
Self-energy (a) and vacuum-polarization (b) diagrams.}
\centerline{ \mbox{
\epsfysize=5cm \epsffile{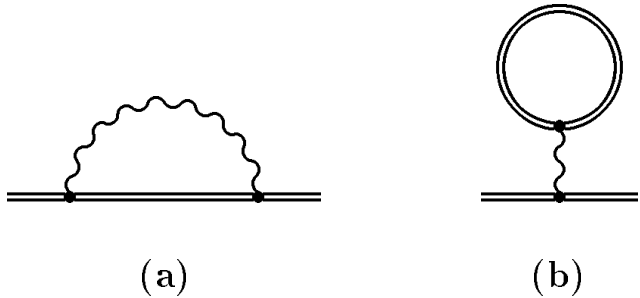}
}}
\end{figure}

Till the beginning of 1980's tests of quantum electrodynamics were mainly
restricted to light atomic systems: hydrogen, helium, 
positronium, and muonium. In these systems, in addition to $\alpha$, 
there is another small parameter, which is $\alpha Z$ ($Z$ is 
the nuclear charge number). For this reason, calculations of light atoms
were based on expansion in these two parameters.
 It means that  with light atomic systems 
we can test QED only to few lowest orders in the parameters
$\alpha$ and $\alpha Z$.
A unique opportunuty to study QED effects to all orders in $\alpha Z$
appeared in experiments with heavy few-electron ions, such as,
for instance, H-like uranium or Li-like uranium. 
High-precision experiments with these ions became possible in the 
last two decades \cite{klu06}.
In heavy few-electron ions the number of electrons is much smaller 
than the nuclear charge number. 
 For this reason,
to the zeroth-order approximation, we can neglect the interelectronic 
interaction and consider that the electrons interact only with the Coulomb
field of the nucleus.
The interelectronic-interaction  and QED effects
 are accounted for by perturbation theory in the
parameters $1/Z$ and $\alpha$, respectively \cite{moh98,sha02}.
For very heavy ions the parameter
$1/Z$ becomes comparable with $\alpha$ and, therefore, 
all contributions can be classified
 by the parameter $\alpha$. 
But, in contrast to light atoms, the parameter $\alpha Z$ is no longer small.
It means that all calculations must be performed without any
expansion in the parameter $\alpha Z$. 

\section{ Binding energy in heavy ions }

\subsection{H-like ions}

To the zeroth-order approximation, a hydrogenlike ion
is described by the Dirac equation with the Coulomb field
of the nucleus ($\hbar=c=1$):
\begin{eqnarray} \label{dirac}
(\balpha \cdot {\bf p}+\beta m+V_{\rm C}(r))\psi({\bf r})=
E\psi({\bf r})\,.
\end{eqnarray}
For the point-charge nucleus, $V_{\rm C}(r)=-\alpha Z/r$,
 this equation can be solved analytically (see, e.g., Refs. \cite{moh98,sha02}). 
To get the binding energy to a higher accuracy 
we need to evaluate quantum electrodynamic
and nuclear corrections.

The finite-nuclear-size correction is evaluated by solving 
the Dirac equation with the potential
induced by an extended nuclear charge-density distribution
and taking the difference 
between the energies for the 
extended and point-charge nucleus. 
This can be done either numerically (see, e.g., Ref. \cite{fra91}) 
or  analytically \cite{sha93}.
In contrast to the nonrelativistic
case, where the corresponding correction 
is completely defined
by the root-mean-square nuclear radius
$\langle r^2\rangle^{1/2}$, in heavy ions the 
higher-order moments of the nuclear charge distribution
may affect the nuclear-size correction on a 1\% accuracy level.

The QED corrections of first order
in $\alpha$ are determined
by the self-energy (SE) and vacuum-polarization (VP) diagrams 
(Fig. 1a,b). 
High-precision calculation
of the SE diagram to all orders in $\alpha Z$
was performed by Mohr \cite{moh74} while
the VP diagram  was first evaluated
by Soff and Mohr \cite{soff88} and by Manakov {\it et al.}
\cite{manakov89}.
 Nowadays calculation of these diagrams causes no problem.


\begin{figure} \caption{Two-loop one-electron Feynman diagrams.}
\centerline{ \mbox{
\epsfxsize=\textwidth \epsffile{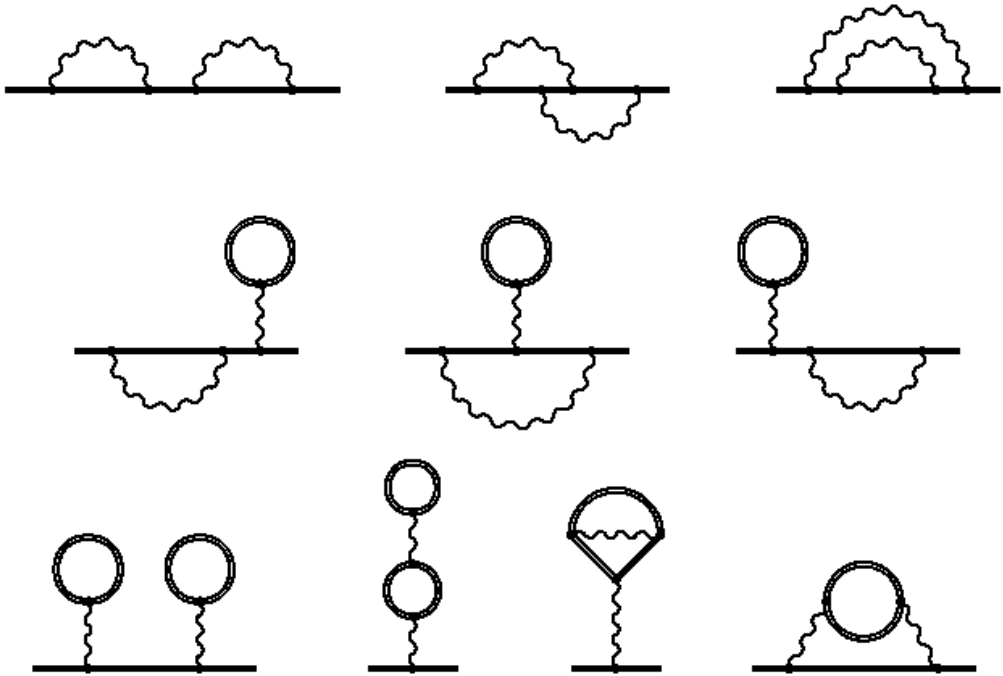}
}}
\end{figure}

The QED corrections of second order in $\alpha$ 
are determined by diagrams depicted in Fig. 2.
Most of these diagrams
can be evaluated by adopting 
the methods developed for the
first-order SE and VP corrections 
\cite{moh98}. 
The most difficult task consists in evaluation
of the SE-SE diagrams  and the combined SE-VP diagrams
presented in the last row of Fig. 2.  
The whole gauge-invariant set 
of the SE-SE diagrams was calculated in
Ref. \cite{yer03}. 
As to the combined
SE-VP diagrams mentioned above, to date they have been 
evaluated only in the free-electron-loop approximation 
(see Ref. \cite{moh98} and references therein).

Next, one should consider the nuclear recoil effect.
It is known that in the nonrelativistic theory of a hydrogenlike
atom the nuclear recoil effect can easily be taken into account 
using the reduced electron mass, $\mu=mM/(m+M)$, where
$M$ is the nuclear mass.
 But this is not the case in the relativistic
theory. The full relativistic theory of the recoil effect
can be formulated only in the framework of QED.
For a hydrogenlike ion, a closed expression 
for the recoil effect to first order in $m/M$
accounting for the complete $\alpha Z$-dependence
 was  derived in Ref. \cite{sha85}
(see also Ref. \cite{sha98a} and references therein). 
Numerical evaluation of 
this expression
to all orders
in $\alpha Z$ was performed in Ref. 
\cite{art95}. 

Finally, one should take into account  
the nuclear polarization correction,
which sets the ultimate accuracy limit  up to which QED can be tested
 with heavy ions. 
This correction is determined by the electron-nucleus interaction
diagrams in which the intermediate nuclear states  are excited. 
It was evaluated in Refs. \cite{plu95,nef96}.

The individual contributions to the ground-state Lamb shift
 in $^{238}{\rm U}^{91+}$, which is defined as the difference
between the exact binding energy and the binding energy derived
from the Dirac equation for the point-charge nucleus,
 are given in Table 1.
The finite-nuclear-size correction is evaluated for the Fermi
model of the nuclear charge distribution with 
$\langle r^2\rangle^{1/2}=5.8507(72)$ fm \cite{ang04}.
The uncertainty of this correction
 is estimated by adding quadratically two errors,
one obtained by varying the root-mean-square radius 
and the other obtained by changing the model of the nuclear-charge
distribution from the Fermi to the homogeneously-charged-sphere
model. 
According to the table,
the present status of the theory and experiment 
on the ground-state Lamb shift in  $^{238}{\rm U}^{91+}$
 provides a test
of QED on the level of about 2\%. 


\begin{table}
\begin{tabular}{lc}
\hline
Finite nuclear size & 198.33(52) \\
QED of first order in $\alpha$ & 266.45 \\
QED  of second order in $\alpha$ & -1.26(33)\\
Nuclear recoil  & 0.46\\
Nuclear polarization & -0.20(10)\\
Total theory &463.78(62)\\
Experiment \cite{gum05} & 460.2(4.6)\\
\hline
\end{tabular}
\caption{Ground-state Lamb shift in $^{238}{\rm U}^{91+}$, in eV.}
\end{table}

\subsection{Li-like ions}

To date, the highest accuracy  
was achieved in experiments 
with heavy Li-like ions \cite{sch91,bra02,bei05}. 
In these systems, in addition to the one-electron 
contributions discussed above,
 one has to evaluate two- and three-electron contributions.
To first order in $\alpha$, the two-electron contribution
is determined by 
the one-photon exchange diagram whose
 calculation causes no problem.
To second order in $\alpha$, one should consider
the two-photon exchange diagrams and
the  self-energy and vacuum-polarization screening diagrams. 
Accurate evaluations of these diagrams were accomplished 
by different groups \cite{art99}.
In addition, to gain the required accuracy,
we need to  evaluate  
the interelectronic-interaction corrections of
third order in the parameter $1/Z$.
The corresponding evaluation within the Breit approximation
was performed in Ref. \cite{zhe00}.

The individual contributions to
the $2p_{1/2}-2s$ transition energy in Li-like uranium are presented 
in Table 2.
The Breit approximation value indicates the transition energy
which can be derived from the Breit equation.
The QED contribution of second order in $\alpha$ 
 incorporates a recent result
for the two-loop SE 
 contribution from Ref. \cite{yer06}. 
The total theoretical value of the transition energy,
$280.76(13)$ eV, 
agrees with 
the related experimental value, $280.645(15)$ eV
 \cite{bei05}.
Comparing the first- and second-order QED contributions
with the total theoretical 
uncertainty, we conclude that
the present status of the  theory and experiment 
for Li-like uranium
 provides a test of QED
on a $0.3$\% level to first order in $\alpha$ and 
on a $8$\% level to second order  in $\alpha$.

\begin{table}
\begin{tabular}{lc}
\hline
Breit approximation & 322.18(9) \\
QED of first order in $\alpha$ & -42.93 \\
QED of second order in $\alpha$ 
& 1.55(9)\\
Nuclear recoil  & -0.07\\
Nuclear polarization & 0.03(1)\\
Total theory &280.76(13)\\
Experiment \cite{bei05}& 280.645(15)\\
\hline
\end{tabular}
\caption{The $2p_{1/2}-2s$ transition energy in Li-like uranium, in eV.}
\end{table}

\section{Hyperfine splitting in heavy ions}

To date, there are several
high-precision measurements of the hyperfine 
splitting (HFS) in heavy hydrogenlike ions~\cite{exp1,exp2,exp4,exp5}.
The hyperfine splitting  of a hydrogenlike ion
can be written as 
\begin{eqnarray}
\Delta E=\Delta E_{\rm Dirac}(1-\varepsilon)+\Delta E_{\rm QED}\,,
\end{eqnarray}
where the Dirac value incorporates the relativistic and nuclear 
charge-distribution effects, $\varepsilon$ is the nuclear magnetization
distribution correction (so-called Bohr-Weisskopf effect), and
$\Delta E_{\rm QED}$ is the QED correction.
The most accurate 
calculations of the QED correction
were performed  in Refs. 
 \cite{sha97,sun98a}.
The theoretical uncertainty is almost completely
determined by the uncertainty of
the Bohr-Weisskopf (BW) effect.
Table 3 presents the ground-state
hyperfine splitting in  $^{209}$Bi$^{82+}$ obtained
by different methods. The theoretical value based
on the single-particle nuclear model \cite{sha97} agrees well with
the experiment \cite{exp1} but has a rather large uncertainty.
The  most elaborated value obtained within a many-particle
nuclear model \cite{sen02} disagrees with the experiment.
A semiempirical evaluation 
employing the experimental value for the HFS in muonic Bi
\cite{eli05} yields the value which deviates by $2\sigma$
from the experiment.
Since
the QED correction is comparable with the uncertainty due to the BW effect,
it is rather
difficult to test QED by the direct comparison of the theory and experiment
on the hyperfine splitting in heavy H-like ions.
However, it has been found that  QED effects on the
HFS can be tested
 by studying a specific difference of the
ground-state HFS values in H- and Li-like ions of the same isotope
 \cite{sha01hfs}. 
Namely, it was shown that the difference
\begin{eqnarray} \label{delprime}
\Delta'E=\Delta E^{(2s)}-\xi \Delta E^{(1s)}\,,
\end{eqnarray}
where $\Delta E^{(1s)}$ and $\Delta E^{(2s)}$ are
 the HFS in H- and Li-like ions of the same isotope,
is very stable with respect to variations of the nuclear 
model, if the parameter $\xi$ is chosen to cancel
the BW corrections in the right-hand side of equation (\ref{delprime}). 
The parameter $\xi$ is almost independent of the nuclear
structure and, therefore, can be calculated to a high accuracy.
In case  of Bi, the calculations yield $\xi$ = 0.16885
and  $\Delta'E$= $-$61.27(4) meV.
The non-QED and QED contributions amount to
$-$61.52(4) meV and 0.24(1) meV, respectively, and, therefore,
 the QED contribution 
is six times larger than the current total theoretical uncertainty.  
This method has a potential to test  QED on 
level of a few percent, provided 
the HFS is measured to accuracy $\sim 10^{-6}$.


\begin{table}
\begin{tabular}{cccc}
\hline 
{Theory  \cite{sha97}}
  & {Theory \cite{sen02}}
  & {Theory \cite{eli05}}
  & {  Experiment \cite{exp1} }   \\
5.101(27)&5.111(-3,+20)(5) & 5.098(7) & 5.0840(8) \\
\hline
\end{tabular}
\caption{Ground-state hyperfine splitting in $^{209}$Bi$^{82+}$, in eV.}
\end{table}

\section{Bound-electron g-factor}

The g-factor of an ion 
 can be defined as the ratio of the magnetic moment 
of the ion to its mechanical moment, expressed in the Bohr
magnetons.
High-precision measurements of the g factor
of H-like carbon \cite{haf00} and oxygen \cite{ver04}
 have triggered a great interest
to related theoretical calculations  
(see Refs. 
  \cite{sha02b,yer02,pac05}
and references therein). In particular, these studies 
provided a new determination of the electron mass
to an accuracy which is four times better than that of the previously
accepted value \cite{moh05}.
An extension of the g-factor investigations
to higher-Z ions could also lead
 to an independent determination of 
the fine structure constant $\alpha$ \cite{wer01}.
The accuracy of such a determination is, however, strongly
limited by a large uncertainty of the nuclear structure
effects which strongly increases when $Z$ is growing \cite{nef02}.
In Ref. \cite{sha06a} it was shown that  the ultimate accuracy limit
can be significantly reduced 
in the difference
\begin{eqnarray} \label{g_dif}
g'=g^{[(1s)^2 (2s)^2 2p_{1/2}]}-\xi g^{[1s]}\,,
\end{eqnarray}
where $g^{[(1s)^2(2s)^2 2p_{1/2}]}$
and $g^{[1s]}$ denote the g-factors of  $^{208}$Pb$^{77+}$
and   $^{208}$Pb$^{81+}$, respectively.
 The parameter $\xi$ must be chosen to cancel the 
nuclear size effect in this difference.
In Ref. \cite{sha06a} 
it was shown
that measurements of the g factor of B- and H-like lead to the same accuracy
as for carbon, accompanied by the corresponding theoretical calculations,
 can provide a determination of $\alpha$ to a higher acuracy
than that from the recent compillation by Mohr and Taylor \cite{moh05}.
This method can also 
 compete in accuracy with the new determination
of $\alpha$ by Gabrielse and co-workers \cite{gab06}.

\section{QED corrections to the $6s-7s$ PNC transition amplitude
in neutral $^{133}$Cs}

The $6s-7s$  PNC transition amplitude in neutral $^{133}$Cs \cite{bou74} 
remains one
of the most attractive subject for tests of the standard model (SM)
at low energies.
The measurement of this amplitude to a $0.3$\% accuracy
\cite{woo97} has stimulated theorists to improve the 
related calculations. 
These improvements, that included evaluations of the Breit 
interaction \cite{der00,koz01},
more precise calculations of the electron-correlation
effects \cite{dzu02}, and calculation of the vacuum-polarization part 
of the QED correction \cite{joh01}, gave
the value for the weak charge of the cesium nucleus, which deviates by 
2$\sigma$ from the prediction of the standard model.
This discrepancy urgently required calculations of the
self-energy part of the QED correction.
Evaluation of the whole gauge-invariant set of 
the SE corrections to the PNC transition amplitude
 with the Dirac-Fock wave functions
was performed in Ref. \cite{sha05}.
The total result for the binding QED correction was found to
amount to $-0.27$\%. This value differs from the previous results 
for the  total binding QED effect,  $-0.5(1)$\% \cite{kuc02} 
and $-0.43(4)$\% \cite{mil02}.
 The discrepency  can be explained
by some approximations made in the previos calculations.
In particular, instead of calculating the QED corrections 
to the PNC amplitude, the previous papers were dealing with 
evaluation of the QED corrections to the PNC mixing coefficient
which is a rather artificial subject for QED. 
However, a semiempirical revision of the previous results 
\cite{fla05} gave a value for the QED correction
which is very close to that of Ref. \cite{sha05} .

Combining the QED correction with other theoretical contributions
and 
comparing the total PNC amplitude with
the experiment for
 an average value for the
vector transition polarizabilty, $\beta=26.99(5)a_{\rm B}^3$
(see Ref. \cite{dzu02} and references therein),
one obtains for the weak charge
of $^{133}$Cs:
\begin{eqnarray}
Q_W=-72.65(29)_{\rm exp}(36)_{\rm th}\,.
\end{eqnarray}
This value
deviates from the SM prediction
 of $-$73.19(13) \cite{ros02}
 by 1.1 $\sigma$.
Further progress on the PNC tests in $^{133}$Cs
can be achieved either by more accurate  atomic
structure calculations or by more precise measurements.

\section*{{Acknowledgements}}
This work was supported in part by RFBR (Grant No. 04-02-17574).
 A.N.A. and D.A.G.
 acknowledge the support by the ``Dynasty'' foundation.

\end{document}